\documentclass[twocolumn,twocolappendix]{aastex631}

\usepackage{amsmath,amssymb,graphicx,booktabs,hyperref}
\usepackage[utf8]{inputenc}
\usepackage[T1]{fontenc}
\usepackage{lmodern}
\hypersetup{colorlinks=true,linkcolor=blue,citecolor=cyan,urlcolor=red}
\shorttitle{A precessing magnetic jet as the engine of GRB 250702B}
\shortauthors{T.~An}

\begin{document}

% ====== METADATA ======
\title{A precessing, magnetically dominated, structured jet powering the hour-scale, quasi-periodic GRB~250702B}

\author[0000-0003-4341-0029]{Tao An}
\affiliation{Shanghai Astronomical Observatory, Chinese Academy of Sciences, 80 Nandan Road, Shanghai 200030, China}

\affiliation{State Key Laboratory of Radio Astronomy and Technology, A20 Datun Road, Chaoyang District, Beijing, P. R. China}

% \date{\vspace{-1.2em}}

\correspondingauthor{Tao An}
\email{antao@shao.ac.cn}

% ====== ABSTRACT ======
\begin{abstract}
GRB 250702B shows ultra-long, episodic prompt activity (three hard $\gamma$-ray episodes over $\simeq 3.2$\,h with quasi-regular spacing $P\simeq 2825$\,s) preceded by a soft X-ray flare about one day earlier. We interpret these phenomena with a unified scenario in which a stellar-mass black hole accretes from a massive, misaligned debris disk and launches a magnetically dominated, precessing, structured (spine–sheath) jet. The engine ``clock'' arises from Lense–Thirring precession of the \emph{outer annulus} of a geometrically thick inner torus at $r \approx 250$--$300\,r_g$, while the hard spectra reflect magnetic-reconnection dissipation in the spine. A slightly off-axis viewing geometry resolves the apparent opening-angle tension without invoking late energy injection. ``Missing'' pulses in the second/third cycles occur naturally when low-amplitude nutation causes the beaming cone to miss the line of sight. The model yields concrete, falsifiable predictions, providing a self-consistent explanation of GRB 250702B’s radiative and outflow anomalies.   
\end{abstract}

\keywords{Gamma‑ray bursts (629) --- Relativistic jets (1390) --- Black hole physics (159) --- Accretion (14) --- Polarimetry (1278)}

% ====== MAIN TEXT ======
\section{Paradox of GRB 250702B}\label{sec1}

On 2025 July 2, a series of high-energy transients, collectively designated GRB 250702BDE, were detected from the same sky location by the Fermi Gamma-ray Burst Monitor (GBM) \citep{GCN40891} and Konus-Wind \citep{GCN40914}. 
The event consisted of three distinct, bright gamma-ray episodes (designated B, D, and E; occurring in chronological order $D\rightarrow B \rightarrow E$), preceded $\approx1$ day earlier by a soft X-ray flare (EP250702a) detected by the \textit{Einstein Probe (EP)} \citep{GCN40906}. Following the literature naming convention, we refer to the event as GRB~250702B.
Multiple high-energy observatories followed the transient (MAXI/GSC, \citealt{GCN40910}; \textit{Swift/XRT}, \citealt{GCN40919}; \textit{NuSTAR}, \citealt{GCN41014}; SVOM/GRM, \citealt{GCN40923}; H.E.S.S, \citealt{GCN41095}; \textit{Chandra}, \citealt{GCN41309}), 
with a second \textit{Chandra} epoch at $\approx65$ d post-burst detecting afterglow decay consistent with $t^{-1.9}$ \citep{OConnor2025Xray}.
The measured temporal separations are $\Delta t$(D$\rightarrow$B) $\approx 2825 \, s$, $\Delta t$(D$\rightarrow$E) $\approx 11551 \, s$, and $\Delta t$(B$\rightarrow$E) $\approx 8727 \, s$, indicating quasi-periodic recurrence with a base period near 2825 s and a later episode consistent with $\approx 4 \times$ this period.  
JWST/NIRSpec spectroscopy has established a secure redshift of $z=1.036\pm0.004$ and localized the transient to an off-nuclear position in a massive, dusty host galaxy \citep{Gompertz2025JWST}.
The near-infrared counterpart was found to be exceptionally red, implying significant dust extinction \citep{GCN41096, GCN40924, 2025arXiv250714286L, Gompertz2025JWST}.

This collection of observational facts presents two profound challenges to our understanding of GRBs. \textbf{(i) A stable, long-period ``engine clock''}. The quasi-regular recurrence of bright $\gamma$-ray pulses with a period of $P \approx 2825$ s ($\approx 47$ minutes) is unprecedented in the known GRB population. Reported quasi-periodic oscillations (QPOs) in GRBs occur on milliseconds to seconds timescales, linked to dynamic processes in accretion disk or central engines \citep{2013ApJ...777..132D, 2022Galax..10...38B, 2023Natur.613..253C, 2025MNRAS.537.2313Y, 2025ApJ...985...33G}. Long-period variability is observed in other accreting systems, such as quasi-periodic eruptions (QPEs) from supermassive black holes (BHs), but these are thermal, lower-luminosity events \citep{2019Natur.573..381M, 2021Natur.592..704A, 2024Natur.634..804N, 2025ApJ...989...13A}, distinct from the non-thermal highly energetic emission of GRB~250702B. \textit{The engine must be both stable over hour-long timescales and capable of repeatedly launching ultra-relativistic outflows.}

\textbf{(ii) A spectral--geometric tension.} The prompt spectrum is unusually hard ($E_{\rm p,rest} \geq 1.2$ MeV) \citep{GCN40914, GCN40931, 2025arXiv250714286L},  
and the event has an extremely large isotropic-equivalent $\gamma$-ray energy $E_{\gamma,\rm iso}\gtrsim 2.2\times10^{54}$ erg \citep{Gompertz2025JWST}, placing it far off the empirical $E_{\rm p}$--$E_{\gamma,\rm iso}$ (Amati) relation \citep{2006MNRAS.372..233A}. 
Moreover, the early-time afterglow was fitted with a uniform ``top-hat'' jet model, implying an ultra narrow jet opening angle of $\theta_{\rm jet} \sim 0.4^\circ$ (\citealt{2025arXiv250714286L}; also Appendix \ref{app:jet}), an order of magnitude smaller than typical long GRBs \citep{2001ApJ...562L..55F, 2003ApJ...594..674B, 2004ApJ...616..331G}, posing a significant challenge for standard jet formation theories \citep{2004IJMPA..19.2385Z, 2005ApJ...629..903L, 2013ApJ...777..162M}. 

  % white dwarf (WD) tidal disruptions by an intermediate-mass black hole (IMBH) \citep{2025arXiv250714286L, EylesFerris2025IMBH}, and ultra-long GRB or sustained engine activity scenarios from comprehensive X-ray observations \citep{2025arXiv250714286L, OConnor2025Xray}, micro-tidal disruption events (micro-TDEs) in which a stellar-mass compact object disrupts a main-sequence star \citep{Beniamini2025microTDE}, a black hole falling into a helium-rich stellar envelope \citep{Neights2025HeStar}. 

% Here we propose a progenitor-independent jet model to interpret this energetic transient at $z = 1.036$: a handful of MeV-hard, $\sim 100 \, s$ pulses repeating quasi-periodically on $\sim$hour-timescales over $\sim3.2\, h$, with faint inter-pulse emission and lingering X-ray activity days later. These observables point to a long-lived, magnetized, structured jet. Once a misaligned BH--disk system launches such a jet, Lense–Thirring precession of the outer collimator, together with a structured (spine--sheath) beam and intermittent magnetic dissipation, naturally reproduces the hour-scale quasi-periodicity, spectral hardness, and ``missing pulses'' while keeping the beaming-corrected energy within the standard bright GRB population. This framework complements progenitor-specific models by providing unified geometric and radiative explanations independent of formation channel.

\section{Progenitor scenarios in literature}\label{sec:channels}

Multiple formation scenarios have been proposed to explain this extraordinary event\footnote{\textit{Context since submission:} five independent studies appeared on arXiv during the review process; they present progenitor scenarios (micro-TDE, BH–into–He-star, WD–IMBH TDE) and multi-wavelength analyses (JWST redshift/energetics; comprehensive X-ray campaign). A detailed discussion is provided in Appendix~\ref{app:comp}.}. These can be broadly grouped into three classes: tidal disruption events (TDEs), stellar collision/merger scenarios, and ultra-long gamma-ray bursts. Here we briefly summarize each channel's key predictions and challenges for  GRB~250702B; detailed comparisons are presented in Appendix~\ref{app:comp}.  

\paragraph{WD--IMBH TDE.}
Disruption of a white dwarf (WD) by an intermediate‑mass black hole (IMBH; $M_{\bullet} \sim 10^{3-4} \, M_\odot$) can power ultra-long high-energy transients with large $E_{\gamma,{\rm iso}}$ and hard spectra \citep{2025arXiv250714286L, EylesFerris2025IMBH}, particularly if a relativistic jet forms. However, TDEs usually occur at galactic nuclei (e.g. \citealt{2021ARA&A..59...21G,2020SSRv..216...32F}), whereas  GRB~250702B is $\sim 1$ kpc off-nucleus. Standard TDE models predict soft, thermal X-ray components from the accretion flow and metal-rich debris signatures (e.g. \citealt{1988Natur.333..523R,1999ApJ...514..180U,2011MNRAS.410..359L,2021ARA&A..59...21G,2020SSRv..216...85S, 1999A&A...343..775K}), which are in tension with the observed MeV-peaked, hard spectrum. Although repeated partial disruptions could in principle produce recurrent outbursts, reproducing the quasi-regular hour-scale base period would require fine-tuned orbits or an additional geometric modulation.

\paragraph{Micro-TDE (stellar‑mass compact object disrupting a star).}
A stellar-mass BH ($M_{\bullet} \sim 3$--$30 \, M_\odot$) or neutron star (NS) disrupting a main-sequence or evolved star \citep{Beniamini2025microTDE} can produce hour-to-day fallback, intermittent fueling, and a relativistic jet. This model can accommodate the long‑lived X‑ray activity. The observed MeV-hard spectrum is feasible if the jet efficiently converts accretion power to radiation. However, the $\sim$47-minute base period favors a geometric clock (e.g., precession) over purely stochastic fallback modulation, suggesting jet dynamics dominate the recurrence regardless of the mass supply rate.

\paragraph{BH into a helium-star envelope.}
A BH spiraling into and accreting from a helium-rich stellar envelope \citep{Neights2025HeStar} can supply sustained accretion over hours to days,  explaining the extreme duration, can provide the required energy reservoir ($\gtrsim 10^{53}$ erg). The dusty, star-forming host environment and slight spatial offset are consistent with massive stellar progenitors. Potential late-time helium-rich spectral signatures in the optical/infrared would be a strong discriminant. As with other scenarios, the stable quasi-periodic modulation points to jet-geometry effects (structured beaming and precession) rather than intrinsic variability in the accretion flow.

\paragraph{Ultra-long GRB / sustained collapsar engines.}
Ultra-long gamma-ray bursts (ULGRBs) powered by collapsars or magnetar \citep{2025arXiv250714286L, OConnor2025Xray} with extended accretion or spin-down can achieve durations of hours and maintain hard MeV spectra. The total energy budget is within the range of the most luminous GRBs. However, reproducing the quasi-periodic $\sim$hour-timescale clock and the ``missing pulses'' requires structured jet geometry (spine-sheath configuration with precession) rather than a simple top-hat jet with an ultra-narrow opening angle $\theta_{\rm jet}$. The off‑nuclear location is unusual but not excluded for massive‑star progenitors in disturbed hosts.

In summary, each scenario supplies a plausible energy source and explains some facets of the data, but \textit{none} predicts the combination of (a) quasi‑periodic hour‑scale pulses, (b) MeV‑hard prompt spectra, and (c) multi‑day X‑ray emission without invoking jet geometry (precession and angular structure). This motivates a \textit{progenitor‑agnostic} focus on jet dynamics and emission geometry, developed in the following section.

\section{A precessing, magnetically-powered, structured jet} \label{sec:mymodel}

\begin{figure}
    \centering
    \includegraphics[width=0.9\linewidth]{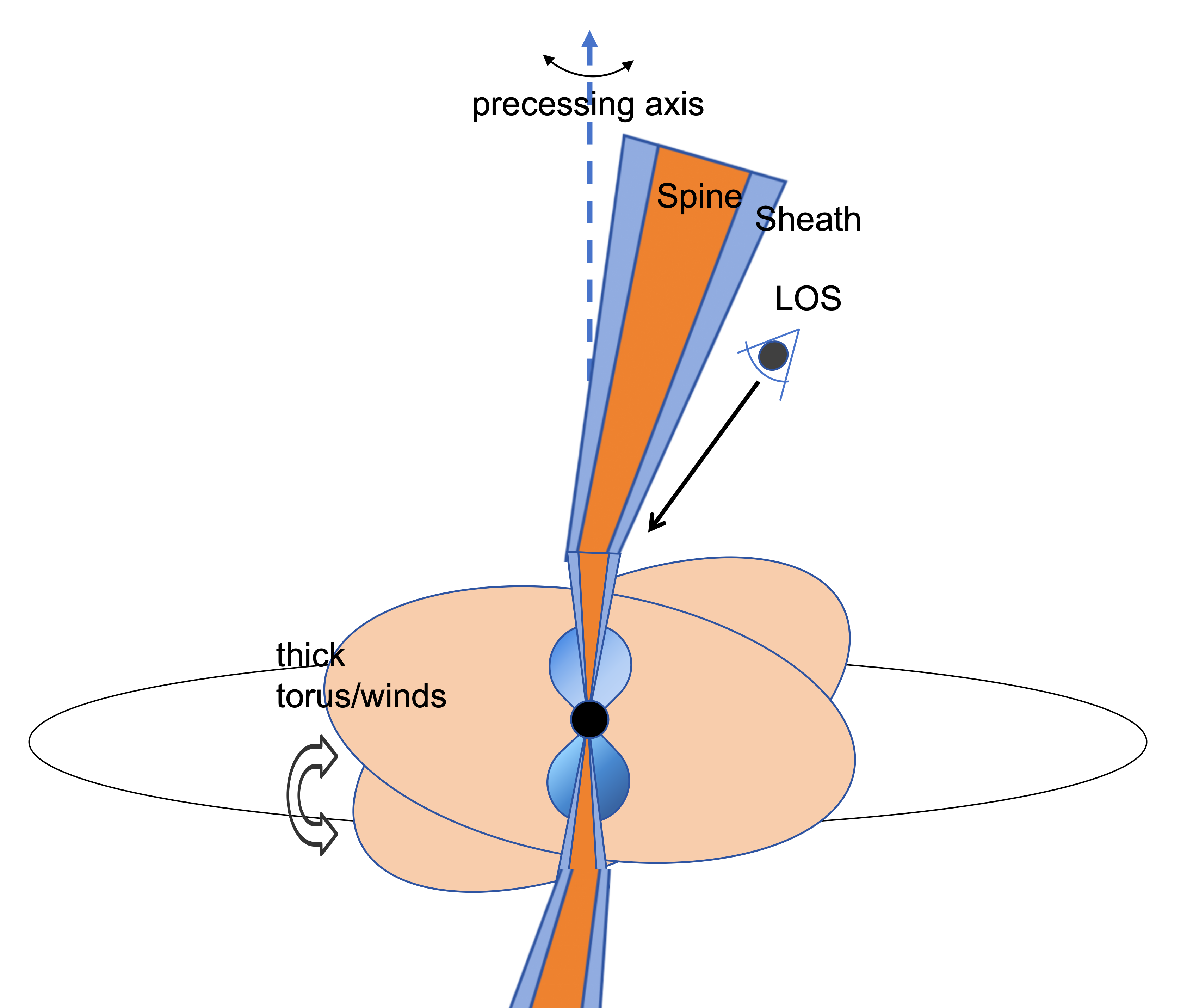}
    \caption{Geometry of the precessing, magnetically dominated structured jet model proposed for GRB 250702B. A misaligned, geometrically thick torus/winds collimates the jet and defines the precessing axis. The structured jet consists of a narrow  spine (orange-color) surrounded by a wider sheath (blue-color), with a small kink at $(2-4) \times 10^2 r_g$ indicating where the jet is forced into precession by the disk wind. The observer’s line of sight (LOS) lies just outside the sheath at a small inclination to the jet axis, enabling lighthouse-like modulation and occasional ``missing'' pulses. An outer thin disk at $r \gtrsim 250\, r_{g}$ is also indicated. }
    \label{fig:sketch}
\end{figure}

We propose a \textit{progenitor-agnostic} unified model in which a long-lived, highly magnetized jet from a misaligned BH–disk system explains all key observables of GRB~250702B. The \textit{clock} is set by Lense–Thirring precession of the outer thick torus/wind that collimates the jet, naturally giving the observed base spacing $P\simeq2825\,\mathrm{s}$ and allowing modest nutation to account for the ``missing'' pulses. The \textit{spectrum} arises from efficient magnetic-reconnection dissipation in a Poynting-flux-dominated flow, producing MeV-peaked, hard prompt emission. The \textit{geometry} is an angularly structured (spine–sheath) jet viewed slightly off-axis (Fig.~\ref{fig:sketch}): precession periodically sweeps the narrow spine across the line of sight (yielding bright cycles) while the wider sheath and afterglow are consistent with a few-degree core rather than an ultra-narrow top-hat, keeping beaming-corrected energies within the bright long-GRB population.

\begin{figure}
  \centering
  \includegraphics[width=0.9\linewidth]{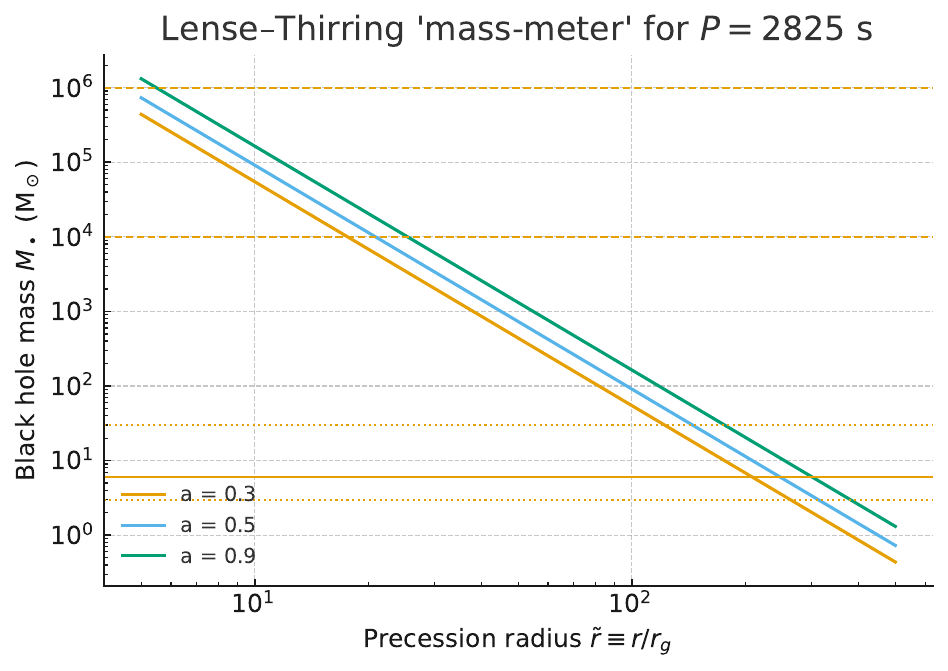}
  \caption{Lense–Thirring ``mass–meter'' for the observed periodicity $P=2825$\,s.
  Curves show the black hole mass $M_\bullet$ implied by a precession radius $\tilde r\equiv r/r_g$ for spins $a_*=\{0.3,0.5,0.9\}$. 
  Horizontal dashed lines mark the IMBH band (10$^{4}$–10$^{6}$\,$M_\odot$);
  dotted lines mark the stellar-mass band (3–30\,$M_\odot$); the thin solid line highlights a GRB-scale $M_\bullet=6\,M_\odot$. For $\tilde r \sim 10 - 50\,r_g$, the period selects the IMBH regime; achieving the same period with a $\sim$6\,$M_\odot$ engine requires $\tilde r\gtrsim 200- 400\,r_g$, or a different precession driver.}
  \label{fig:LT_mass_meter}
\end{figure}

\paragraph{The Engine Clock: Lense-Thirring Precession} The quasi-regular period of $P \approx 2825$ s is identified with the Lense-Thirring (LT) precession \citep[e.g.,][]{1972ApJ...178..347B} of the geometrically thick inner flow --- specifically, the \emph{outer working annulus} of the thick torus/wind that collimates the jet around a spinning black hole \citep{1998ApJ...492L..59S,2009MNRAS.397L.101I, 2025ApJ...980..119B}. For a ring at radius $r$:
    \begin{equation}
    P_{\rm LT} = \frac{2\pi}{\Omega_{\rm LT}} \approx \frac{\pi G M_{\bullet}}{a_* c^3} \left( \frac{r}{r_{\rm g}} \right)^3,
    \end{equation}
    where $a_*$ is the dimensionless black hole spin parameter and $r_{\rm g} \equiv G M_{\bullet} / c^2$ is the gravitational radius. For $M_{\bullet} = 6 M_\odot$ and a moderate spin of $a_* \sim 0.5-0.9$, the observed period $P \approx 2825 \, s$ implies $r \sim 250-300 \, r_{\rm g}$ (Fig.~\ref{fig:LT_mass_meter}). 
    It is convenient to express this in the Kerr scale 
    following \citet{2024ApJ...972L..23V}: $K \equiv r/r_{\rm ISCO}(a_\ast)$ for prograde orbits.
    % \begin{equation}
    %     K \equiv \frac{r}{r_{\rm ISCO}(a_*)}, \quad r_{\rm ISCO}(a_*)~\text{for prograde orbits}.
    % \end{equation}
    For a Kerr black hole, this gives $r_{\rm ISCO}/r_g=\{4.233,\,3.393,\,2.321\}$ for $a_*=\{0.5,\,0.7,\,0.9\}$, hence, we get $K=\{59,\,82,\,130\}$. 
    The large $K$ values indicate that the \emph{period is set by the outer working annulus of the inner thick torus/wind (the jet collimator)}, rather than by the innermost nozzle near $r_{\rm ISCO}$. In a super-Eddington, geometrically thick state this annulus can precess quasi-rigidly, providing a stable engine ``clock'' over multiple cycles (Appendix~\ref{app:detailedmodel}). By contrast, the innermost flow is partially aligned by the Bardeen–Petterson effect and does not sweep across the line of sight; residual torques from a misaligned \emph{outer thin disk} excite only low-amplitude nutation of the precession axis, intermittently suppressing pulses without invoking multiple engines.    

\paragraph{The Energy Source: Magnetic Reconnection}
    A highly magnetized, Poynting-flux-dominated jet (magnetization parameter $\sigma \gg 1$) dissipates via relativistic magnetic reconnection, efficiently accelerating particles and producing hard synchrotron spectra \citep{2005MNRAS.358..113L,2010MNRAS.402..353L,2010MNRAS.408L..46G,2011ApJ...726...90Z,2014ApJ...783L..21S,2014PhRvL.113o5005G,2016ApJ...816L...8W}.  The \emph{observer-frame} peak energy is given by:
    \begin{equation}
    E_{p,{\rm obs}}\ \sim\ \frac{3 h q_e B'\,\gamma_{e,\mathrm{peak}}^2\,\Gamma}{4\pi m_e c\, (1+z)}\,,\qquad E_{p,{\rm rest}}=(1+z)E_{p,{\rm obs}},
    \end{equation}
    with bulk Lorentz factor $\Gamma$, comoving magnetic field strength $B'$, and characteristic electron energy $\gamma_{e,\mathrm{peak}}$ \citep{1998ApJ...497L..17S}.  For GRB~250702B ($z=1.036$), a fiducial set
    $\Gamma\simeq 250 - 300$, $B'\simeq(8 - 10)\times10^3$ G, $\gamma_{e,\mathrm{peak}}\simeq(6 - 8)\times10^3$, and $\sigma\sim20 - 50$
    yields $E_{p,{\rm rest}}\simeq 1 - 3$ MeV with high radiative efficiency, consistent with the observations.

\paragraph{The Geometry: A Structured Jet} The outflow is angularly structured rather than a uniform ``top-hat'': a narrow ultra-relativistic core/spine ($\theta_c \sim 3^\circ$--$5^\circ$) surrounded by a wider sheath. A slightly off‑axis viewing ($\theta_v \gtrsim \theta_c$)  reproduces the early rise and apparent ``early break'' (Appendix \ref{app:jet}). Top-hat fits can spuriously return an ultra-narrow $\theta_{\rm jet}$, which we interpret as a geometric illusion. Such a spine--sheath structure is a generic outcome of jet propagation through the dense environment surrounding a collapsar or merger remnant \citep{2004IJMPA..19.2385Z,2013ApJ...777..162M,2018PhRvL.120x1103L}.

% A two-component (spine–sheath) jet is likewise a generic outcome of jet propagation in dense environments and structured outflows \citep[e.g.][]{2002MNRAS.332..945R,2004IJMPA..19.2385Z,2013ApJ...777..162M,2018PhRvL.120x1103L}, and has also been discussed in the GRB context by \citet{1999Sci...284..115V,2001ApJ...552L..31V}. 
% For GRB~250702B, $q\!\sim\!10^{-4}$ naturally places the hours-long duration between canonical long GRBs ($q\!\sim\!10^{-2}$) and SLSNe ($q\!\sim\!10^{-7}$), as shown in Fig.~\ref{fig:K-q}.

\begin{figure}
    \centering
    \includegraphics[width=0.9\linewidth]{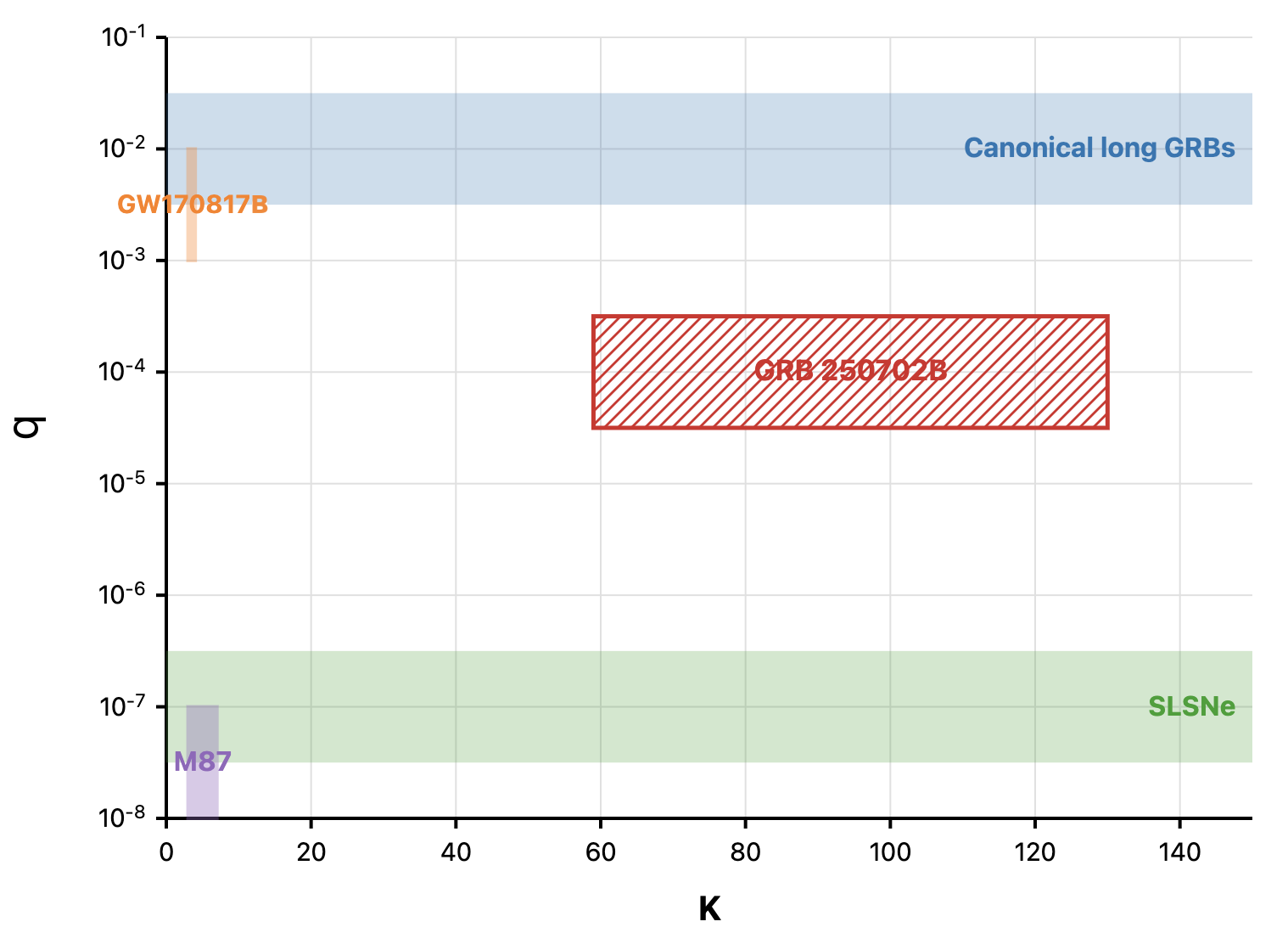}
    \caption{Qualitative placement of GRB~250702B (shaded rectangle) in $(K, q)$ space, where $K \equiv r/r_\mathrm{ISCO}(a_*)$ and $q \equiv M_\mathrm{T}/M_\mathrm{BH}$ is the torus-to-BH mass ratio \citep{2024ApJ...972L..23V}. For the observed period $P \simeq 2825$\,s and a fiducial $M_\mathrm{BH} = 6\,M_\odot$, GRB~250702B occupies $K \simeq 59$--$130$ for BH spins $a_* = 0.5$--$0.9$ and $q \sim 10^{-4}$ (shaded band). The adopted $q$ value places the event between canonical long GRBs ($q \sim 10^{-2}$) and SLSNe ($q \sim 10^{-7}$). Rectangular boxes show representative $(K, q)$ regions for GW170817B (the post-merger black-hole spin-down signal associated with GW170817 with estimated $K\sim 3-4$ and $q \sim 10^{-3}$--$10^{-2}$, \citealt{2024ApJ...972L..23V}) and M~87 with minimal torus fraction $q \sim 10^{-7}$--$10^{-8}$ (jet-launching and collimation occur on scales of a few $r_{\rm ISCO}$, inferred from VLBI constraints and jet kinematics, e.g., \citealt{2012Sci...338..355D, 2016A&A...595A..54M, 2018A&A...616A.188K} and further supported by the detection of jet precession e.g., \citealt{2023Natur.621..711C}).  Both GW170817B and M\,87 therefore anchor the ``few--$r_{\rm ISCO}$'' end of the diagram, whereas GRB~250702B requires a much larger $K$ where the precession is set by the outer working annulus of a thick inner flow. }
    \label{fig:K-q}
\end{figure}

\paragraph{Scale-free duration ordering (q-scaling)}
    We define $q  \equiv M_{\rm T}/M_{\bullet}$ (torus-to-BH mass ratio)\footnote{Some papers denote the torus-to-BH mass ratio by $\sigma$; to avoid confusion with the jet magnetization in this study, we use $q$ for the mass ratio and $\sigma$ for magnetization.}. 
    When the prompt duration is set by the longevity of the inner working torus, a supply-limited scaling $T \sim t_{\rm visc} \propto M_{\rm T}/\dot M$ at fixed $M_{\bullet}$ implies the simple ordering $T \propto q^{-1}$ \citep[for the original cross-mass formulation,][]{2003ApJ...584..937V,2017MNRAS.464.3219V}. 
    Here $q$ refers to the mass of the inner working torus that sets longevity; the extended misaligned disk at large $K$ acts as collimator/reservoir and need not be small. 
    This interpretation is consistent with the viscous/warp physics and coherent precession of geometrically thick, super-Eddington tori, independently of the detailed power-extraction mechanism \citep[e.g.][]{1983MNRAS.202.1181P,2007ApJ...668..417F,2009MNRAS.397L.101I,2014MNRAS.439..503S,2023ApJ...955...72K}.
    A two-component (spine--sheath) jet is likewise a generic outcome of jet propagation in dense environments and structured outflows \citep[e.g.][]{2002MNRAS.332..945R,2004IJMPA..19.2385Z,2013ApJ...777..162M,2018PhRvL.120x1103L}, and has also been discussed in the GRB context by \citet{1999Sci...284..115V,2001ApJ...552L..31V}.  
    For  GRB~250702B, $q\sim10^{-4}$ naturally places the hours-long duration between canonical long GRBs ($q \sim 10^{-2}$) and SLSNe ($q \sim10^{-7}$) \citep{2017MNRAS.464.3219V}, shown in Fig.~\ref{fig:K-q}.

\paragraph{Duty cycle and power}
The prompt emission comprises three $\sim100$\,s pulses over $\sim3.2$\,hr, yielding a duty cycle $\delta\simeq(100\times3)/11520\approx0.026$. Adopting the JWST prior $E_{\gamma,\mathrm{iso}}\gtrsim2.2\times10^{54}$\,erg, and for a structured Gaussian jet with core width $\theta_c=3^\circ$–$5^\circ$ and a modest off-axis view $\theta_v\simeq6^\circ$–$8^\circ$, the \emph{intrinsic core energy} (corrected for jet structure but not for viewing angle) is $E_{\gamma,\mathrm{core}}\approx(3$–$8)\times10^{51}$\,erg, placing GRB\,250702B within the bright long-GRB population; the \emph{observed} energy for $\theta_v>\theta_c$ is smaller due to Doppler de-boosting.
The three $\approx100$\,s bright windows within $\approx3.2$\,hr are set by geometric visibility of a precessing, structured jet rather than steady isotropic emission. Hence rescaling the total energy by $1/\delta$ is not warranted: the measured $E_{\gamma,\mathrm{iso}}$ already accounts for direction-dependent beaming during the bright phases. A beaming-corrected estimate based on the Gaussian core yields $E_{\gamma,\mathrm{core}}\approx(3$–$8)\times10^{51}$\,erg and a time-averaged power over $3.2$\,hr of $\langle L_\gamma\rangle\approx(3$–$7)\times10^{47}$\,erg\,s$^{-1}$. The $\sim100$\,s pulse widths follow from the finite visibility window of an off-axis, structured jet with $\theta_{\rm eff}\!\simeq\!\theta_c+1/\Gamma$ sweeping at $\Omega_{\rm LT}$; for representative parameters this gives $\Delta t_{\rm vis}\!\simeq\!80$–$140$\,s rather than the $\sim16$\,s expected for a $1^\circ$ top-hat (see Appendix~\ref{app:geometry} for the analytic estimate and geometry).

\section{Physical interpretation of radiative and outflow anomalies} \label{sec:phys}

We interpret the ensemble of observational anomalies in GRB~250702B within a single, magnetically dominated, precessing-jet framework that self-consistently links the prompt temporal structure, high radiative efficiency, and post-prompt afterglow behavior. A compact remnant surrounded by a massive, misaligned debris disk \citep{1999ApJ...524..262M,2008MNRAS.388.1729K,2007ApJ...668..417F}  drives super-Eddington accretion that inflates a geometrically thick inner flow \citep{1999ApJ...524..262M, 2008MNRAS.388.1729K, 2014MNRAS.439..503S}. The \emph{outer working annulus} of this inner torus, located at $r \sim K\,r_{\rm ISCO}$ with $K \gg 1$, undergoes quasi-rigid Lense–Thirring precession \citep{2007ApJ...668..417F, 2009MNRAS.397L.101I} and thus sets the observed $\sim$kilosecond clock. The same accretion state favors a Poynting-flux–dominated jet launched via the Blandford–Znajek mechanism \citep{1977MNRAS.179..433B,2011MNRAS.418L..79T}, whose dissipation proceeds primarily via magnetic reconnection in a structured spine–sheath outflow \citep{2011ApJ...726...90Z}. Small outer-disk torques are expected to induce low-amplitude nutation; details are discussed in Section \ref{sec:mymodel}.

% This unified framework provides a   physical sequence that explains all the peculiar features of GRB~250702B. A collapsar or compact object merger produces a stellar-mass black hole surrounded by a massive, misaligned debris disk \citep{1999ApJ...524..262M,2008MNRAS.388.1729K,2007ApJ...668..417F}, naturally associated with star-forming environments and  consistent with the off-nuclear location. The high fallback rate drives super-Eddington accretion, inflating the inner flow into a thick torus \citep{1999ApJ...524..262M, 2008MNRAS.388.1729K, 2014MNRAS.439..503S} that is susceptible to quasi-rigid LT precession, setting the clock \citep{2007ApJ...668..417F, 2009MNRAS.397L.101I}. This same accretion state is the ideal environment for launching a Poynting-flux-dominated jet via the Blandford-Znajek mechanism \citep{1977MNRAS.179..433B,2011MNRAS.418L..79T}. The dissipation of this magnetic energy via reconnection powers the hard prompt emission \citep{2011ApJ...726...90Z}, while jet-ambient interaction sculpts the angular structure \citep{2013ApJ...777..162M,2018PhRvL.120x1103L}.

\paragraph{Geometric gating of prompt emission}
The observed gamma-ray pulses arise through geometric gating: the precessing, narrow jet spine sweeps across the observer's line of sight, which is positioned at a viewing angle $\theta_v > \theta_c$. Bright flashes occur only when the spine's beaming cone (half-angle $\approx 1/\Gamma$) intersects our line of sight. For $\Gamma\sim100$--$300$, this corresponds to $\approx 0.2^\circ$--$0.6^\circ$. The missing pulses between episodes B and E, as well as the near-integer multiple spacings, arise from modest nutation of the precession axis \citep{2007ApJ...668..417F,2018MNRAS.474L..81L}, which causes the jet spine to narrowly miss the line of sight during some cycles. Such nutation is likely driven by torques from the misaligned outer accretion disk. The unified treatment of the multiple GBM triggers as a single source exhibiting episodic activity over $\sim 10$--$30$~ks, together with the characteristic inter-episode timescale of $\sim2.8$~ks in the observer frame, strengthens this geometric-gating hypothesis. In a thick, quasi-rigidly precessing torus, the LT frequency scales as $\Omega_{\rm LT}\propto a_\ast (r/r_g)^{-3}$ (mass-weighted across the flow); the observed spacing then provides a direct constraint on the precessing annulus. Small nutations can account for skipped sightings without requiring separate engines or distinct explosion sites.

\paragraph{Radiative channels, efficiency, and spectral behavior}
A high-$q$ (Poynting-dominated) spine embedded in a slower shear layer explains both the high radiative efficiency and the pulse-level spectral hardness: a magnetic-reconnection-fed ``mini-jet'' intermittently beams into (or just past) the line of sight, while shear regulates pair loading and stabilizes the comoving compactness. The intermittency is therefore primarily geometric rather than driven by stochastic dissipation-rate variability, which also explains the hours-long duty cycle coordinated across \textit{EP} and \textit{Fermi}/GBM bands \citep{OConnor2025Xray}. This picture avoids fine-tuned photospheric parameters and is consistent with the absence of strong high-energy ($\gtrsim$MeV) spectral components during the late GBM episodes.

\paragraph{Afterglow evolution and jet structure}
This geometry resolves the opening-angle tension. The off-axis afterglow of a structured jet exhibits a characteristic rising light curve that peaks when the jet's bulk Lorentz factor decelerates to the point where $1/\Gamma \sim (\theta_v - \theta_c)$ \citep[e.g.,][]{2002ApJ...570L..61G,2002MNRAS.332..945R}. Misidentifying this peak as an early on-axis jet break would lead to a severe underestimation of the true jet opening angle. The inferred $\theta_{\rm jet} \sim 0.4^\circ$ is thus likely a modeling artifact; our framework accommodates a core width of $\theta_c \sim 3^\circ$--$5^\circ$ consistent with the data (Appendix \ref{app:jet}). 

The new \textit{Chandra} epochs at $\approx38$ and $\approx65$~d (observer frame) detect a fading X-ray source precisely at the NIR/radio position with a temporal slope $\alpha_X\simeq1.8$--$1.9$ referenced to the initial GBM trigger, and no late rebrightening. The X-ray decay is consistent with a post-jet-break external shock from a slightly off-axis Gaussian jet (e.g., $\theta_c\sim3^\circ$--$5^\circ$, $\theta_v\sim6^\circ$--$8^\circ$) and disfavors substantial late energy injection. Contemporaneous radio detections (VLA at 10~GHz \citealt{GCN41053}; MeerKAT and uGMRT at 1--3~GHz \citealt{GCN41147, GCN41145}) are well within the forward-shock parameter space implied by the X-rays, providing a   broadband afterglow with no need for refreshed shocks. A joint \textit{Swift}--\textit{NuSTAR}--\textit{Chandra} analysis likewise prefers a single external-shock component with no chromatic anomalies. The off-axis peak time $t_{\rm peak}$ and the post-break slope jointly constrain the degeneracy among $(\theta_c,\theta_v,\epsilon_e,\epsilon_B,n)$: for the allowed microphysics priors, $t_{\rm peak} \sim 10^4$~s cannot be reproduced with $\theta_{\rm jet}\lesssim1^\circ$ if the late \textit{Chandra} slope is $|\alpha_X|\gtrsim1.8$. The absence of plateaus or rebrightenings in 0.3--10~keV at 38--65~d and the consistent radio--X-ray SED remove the need for late-time energy injection; any residual deviations would require $E_{\rm inj}\ll E_{\rm K,iso}$ and would be sub-dominant.  

\paragraph{Host environment and progenitor constraints}
At $z\simeq1.036$, the isotropic-equivalent energetics inferred for the prompt and afterglow emission securely place the event in the GRB regime unless unusually large beaming corrections apply to an alternative channel. The heavy line-of-sight extinction and the non-nuclear offset \citep{Gompertz2025JWST} are both typical of long GRBs in dusty disks, while simultaneously compatible with a jetted TDE scenario. A targeted re-evaluation of the IMBH--WD TDE hypothesis shows that, given the measured redshift and X-ray luminosity, any TDE interpretation must be strongly relativistic (jetted) and hence observationally difficult to distinguish from a GRB purely on afterglow grounds; however, the absence of an additional thermal optical/NIR component at rest-frame $\sim$2--6~weeks mildly disfavors a luminous, slowly cooling disk/reprocessing layer. With $z\simeq1.036$, a non-relativistic TDE is excluded on Eddington grounds. A jetted TDE remains viable only if a narrowly beamed jet replicates standard GRB afterglow scalings, in which case our jet-based interpretation still applies to the radiation physics and angular structure.

\section{Predictions and Future Tests}

Our model makes several specific and falsifiable predictions.

% Final minimal-change version of the closing enumerated predictions for §4. % Only wording refinements or compact additions are made; substantive meaning preserved.
% (1) \emph{Polarization tomography:} IXPE-like measurements during bright intervals should show phase-coherent polarization-angle swings at the precession period; the degree should peak when the beaming cone grazes the line of sight. (2) \emph{Radio size and scintillation:} Early-time diffractive/weak-refractive scintillation followed by quenching at $t\sim{\rm weeks}$ will constrain the angular core and test the inferred $(\theta_c,\theta_v)$ independent of light-curve modeling; VLBI non-detection of superluminal centroid motion would be consistent with a modest off-axis geometry. (3) \emph{Late X-rays:} Continued \textit{Chandra}/\textit{XMM} monitoring should follow a single power law with no chromatic breaks; any strong late-time spectral softening or rebrightening would argue for sustained central-engine power or dense-shell CSM interaction. (4) \emph{Spectral--temporal closure:} With the measured $z$ and $A_{V}$, multi-band radio--X-ray closure relations predict $\beta_X\approx(p/2)$ and $\alpha\approx(3p-2)/4$ post-break; deviations would point to evolving microphysics or angular-energy stratification beyond a simple Gaussian.

\begin{enumerate}
    \item \textbf{Phase-resolved polarization:} We predict periodic swings in the polarization angle and modulated linear polarization as the jet spine, with its ordered magnetic field, precesses across our line of sight  \citep{2003ApJ...597..998L,2009ApJ...698.1042T,2018MNRAS.474L..81L}. These variations should be phased to the gamma-ray pulses.
    
    \item \textbf{Fourier comb in the gamma-Ray signal:} Time-series analysis of the gamma-ray light curve should reveal strong power at the fundamental precession frequency $f_0 = 1/P_{\rm LT} \approx 3.5 \times 10^{-4}$ Hz and its harmonics ($2f_0, 3f_0, \ldots$). This harmonic structure should persist even during the quiescent intervals between bright pulses. 
    
    \item \textbf{Late-Time afterglow:} 
    The new \textit{Chandra} detection at $T_0\approx65$ days shows a smooth, steep X-ray decline ($F_X \propto t^{-1.9}$) with no plateau/rebrightening \citep{OConnor2025Xray}. 
    Continued \textit{Chandra}/\textit{XMM} monitoring should follow a single power law without chromatic breaks; any late spectral softening or rebrightening would argue for sustained central-engine power or dense-shell CSM interaction.

    \item \textbf{Host constraints from JWST observations:} Published JWST spectroscopy and imaging reveal a dusty, star-forming host with a non-nuclear offset, consistent with our model. Remaining tests include: (i) search for persistent narrow emission lines or a thermal NIR component at late times signaling disk reprocessing; (ii) verification that host star formation rate and metallicity lie within the long-GRB locus while remaining compatible with a jetted-TDE alternative. Non-detections would favor a jet-dominated, off-axis afterglow without substantial disk contribution.    

    \item \textbf{Radio size and morphology:} Very Long Baseline Interferometry imaging at $t\gtrsim$months can directly measure the angular size and constrain $(\theta_c,\theta_v)$ independent of light-curve modeling. Non-detection of superluminal centroid motion would be consistent with a modest off-axis geometry, while a resolved asymmetric morphology could reveal jet stratification.
\end{enumerate}

Incorporating the EP pre-trigger activity, the consolidated GBM interpretation, the JWST redshift/host diagnostics, and the \textit{Chandra} late-time decay, the simplest self-consistent picture is a magnetically dominated, precessing, structured jet viewed slightly off-axis. This geometry explains the ultra-long, intermittent prompt behavior and removes the opening-angle tension without late energy injection, while remaining agnostic about the progenitor (collapsar vs.\ jetted TDE). At the same time, geometric gating is generic to any precessing, structured jet, implying a progenitor degeneracy. Discriminating among channels therefore requires late-time diagnostics beyond jet geometry, e.g., He-rich emission (BH$\rightarrow$He-star), supernova signatures (collapsar/ULGRB), or neither (micro-TDE), together with polarization monitoring and late-time radio/X-ray calorimetry.

\appendix

% \twocolumngrid   % <-- if you want to keep 2-column
% OR
\onecolumngrid   % <-- switch to 1-column mode

\section{Detailed Model Calculations and Derivations} \label{app:detailedmodel}

\subsection{Lense-Thirring Precession of a Thick Torus} \label{app:LT}

The Lense-Thirring (LT) effect arises from  frame-dragging in the spacetime of a rotating mass, as described by the Kerr metric \citep{1918PhyZ...19..156L, 1972ApJ...178..347B}. For a test particle in a circular orbit of radius $r$ inclined with respect to the black hole's equatorial plane, the orbital plane precesses with an angular frequency:
\begin{equation}
    \Omega_{\rm LT} = \frac{2 G J}{c^2 r^3} = \frac{2 a_* G^2 M_{BH}^2}{c^3 r^3},
\end{equation}
where $J = a_* G M_{\bullet}^2 / c$ is the angular momentum of the black hole and $a_*$ is the dimensionless spin parameter. Expressing $r$ in units of the gravitational radius $r_g = G M_{\bullet} / c^2$, so $r = \tilde{r} r_g$. The precession frequency becomes:
\begin{equation}
\Omega_{\rm LT} = \frac{2 a_* c}{\tilde{r}^3 r_g} = \frac{2 a_* c^3}{G M_{\bullet} \tilde{r}^3 } \end{equation}
The corresponding precession period, $P_{\rm LT} = 2\pi / \Omega_{\rm LT}$, is therefore:
\begin{equation}
P_{\rm LT} = \frac{\pi G M_{\bullet}}{a_* c^3} \tilde{r}^3, \, \tilde{r} = \left( \frac{P_{\rm LT} a_* c^3}{\pi G M_{\bullet}} \right)^{1/3} = 246 \left( \frac{a_*}{0.5} \right)^{1/3} \left( \frac{M_{\bullet}}{6 M_\odot} \right)^{1/3} \left( \frac{P_{\rm LT}}{2825\,{\rm s}} \right)^{1/3} \end{equation}
as used in the main text.
For GRB~250702B ($P_{\rm LT} = 2825$ s, $M_{\bullet} = 6 M_\odot$), this yields $\tilde{r} \approx 250$ for $a_* = 0.5$ and $\tilde{r}  \approx 300$ for a higher spin $a_* = 0.9$. These radii, corresponding to $\sim 2.2 \times 10^8$ cm and $\sim 2.6 \times 10^8$ cm respectively. These values lie well within the photospheric radius $r_{\rm ph} \sim 10^9$ cm inferred from thermal emission modeling and are consistent with massive super-Eddington accretion tori in collapsar models \citep{1999ApJ...524..262M, 2011MNRAS.410.2385T, 2020MNRAS.499.4097Z, 2022MNRAS.510.4962G}.\textbf{This makes Lense-Thirring precession a physically plausible clock mechanism in this context.} The weak scaling $\tilde{r} \propto M_{\bullet}^{1/3} a_*^{1/3}$ yields $\pm 6\%$ variation across $5-7.5\,M_\odot$ and $\pm 15\%$ across $0.3 \leq a_* \leq 0.95$.

\textbf{Conditions for Coherent Rigid-Body Precession.}
For an accretion disk to precess as a rigid body, the internal communication timescale must be much shorter than $P_{\rm LT}$. A standard, geometrically thin, viscous accretion disk would be subject to strong differential precession, and will warp and tear apart \citep{1975ApJ...195L..65B,2012MNRAS.421.1201N}. In contrast, a geometrically-thick, pressure-supported torus can precess quasi-rigidly if the this condition is met \citep{1983MNRAS.202.1181P,2007ApJ...668..417F,2009MNRAS.397L.101I}.

The sound-crossing time across the torus, $t_s \sim r / c_s$ (where $r$ is the torus radius and $c_s$ is the sound speed), must satisfy $t_s \ll P_{\rm LT}$. For a radiation-pressure-dominated flow, characteristic of super-Eddington accretion, the sound speed is $c_s \approx \sqrt{4 P_{\rm rad} / (3 \rho)} \approx c / \sqrt{3}$; thus the sound crossing time is $t_s \sim r / c_s \sim \sqrt{3} r / c$. At a radius of $r = 250 r_g$ for a $6 M_\odot$ black hole ($r \approx 2.2 \times 10^8$ cm), $t_s \sim 1.3 \times 10^{-2}$ s, which is orders of magnitude shorter than the observed period $P_{\rm LT} \approx 2825$~s. This confirms that the inner flow can precess coherently.

To assess precession stability, we require (i) rigid-body coupling, (ii) rapid warp communication relative to the LT period, and
(iii) slow radial inflow compared to the LT period. In the bending-wave regime
(\(\alpha \lesssim H/r\)), the warp communication time
\(t_{\rm wave}\sim 2(r/H)\,\Omega_K^{-1}\) satisfies
$
t_{\rm wave}/P_{\rm LT}
= \frac{2}{\pi}\,a_*\, (r/H) (r_g/r)^{3/2}\ll 1
\quad\text{at } r\simeq 250\,r_g\ \ (\text{for } a_*\sim 0.5,\ H/r\sim 0.1),
$
ensuring coherent precession.  
Consequently, the  observed hour-scale duration $T$ (in Section 3 `Scale-free duration ordering') can be understood as the outcome of a small inner $q$ governing the engine lifetime, together with a thick, low-$\alpha$ outer collimator at large $K$ whose bending-wave communication enables multi-cycle coherent precession.

The viscous time
\(t_{\rm visc}=r^2/\nu=(1/\alpha)(r/H)^2\,\Omega_K^{-1}\) then controls longevity:
\begin{equation}
\frac{t_{\rm visc}}{P_{\rm LT}}
= \frac{a_*}{\pi\,\alpha}\left(\frac{r}{H}\right)^{2}\left(\frac{r_g}{r}\right)^{3/2}  \approx 2.0 \left( \frac{10^{-3}}{\alpha} \right) \left( \frac{0.15}{H/r} \right)^2 \left( \frac{\tilde{r}}{250} \right)^{-3/2}.
\end{equation}
Evaluated at a baseline disk thickness of $H/r = 0.15$, $r \simeq 250\,r_g$ and $\alpha = 10^{-3}$, we find $t_{\rm visc}/P_{\rm LT} \approx 2.0$, indicating marginal coherence. 

In geometrically thick, low-$\alpha$ disks the warp propagates as \emph{bending waves} when $\alpha < H/r$, so the key condition for quasi-rigid precession is \emph{rapid warp communication} ($t_{\rm wave} \ll P_{\rm LT}$) rather than viscous diffusion. For our baseline parameters ($H/r=0.15$, $\tilde r \equiv r/r_g\simeq250$–300), this condition is satisfied; adopting $\alpha=10^{-3}$ further yields $t_{\rm visc}/P_{\rm LT} \approx 2$, ensuring multi-cycle coherence ($t_{\rm visc} \gtrsim P_{\rm LT}$) in addition to $t_{\rm wave} \ll P_{\rm LT}$. While classical magnetorotational instability turbulence often gives $\alpha \sim 0.01$–0.1 \citep{1995ApJ...440..742H}, super-Eddington, radiation-dominated flows can exhibit reduced effective viscosity \citep{2009MNRAS.397L.101I,2021MNRAS.507..983L}, with recent GRMHD results finding $\alpha_{\rm eff} \sim 10^{-3}$–$10^{-2}$ \citep{2023ApJ...955...72K}, consistent with the values required here.

\subsection{Particle Acceleration and Spectral Formation via Magnetic Reconnection} \label{app:MagRec}

The hard prompt spectrum of GRB~250702B, with rest-frame peak energy $E_{p,{\rm rest}} \gtrsim 1.2$ MeV and high radiative efficiency, requires efficient non-thermal particle acceleration beyond standard shock models. In Poynting-flux-dominated jets with magnetization parameter $\sigma \gg 1$, relativistic magnetic reconnection provides the primary dissipation mechanism, efficiently converting magnetic energy into accelerated particles and hard synchrotron emission \citep{2005MNRAS.358..113L, 2010MNRAS.408L..46G, 2011ApJ...726...90Z}.

Large-scale particle-in-cell (PIC) simulations show this process  efficiently converts a significant fraction of the magnetic energy into a non-thermal particle population \citep{2014ApJ...783L..21S,2014PhRvL.113o5005G,2016ApJ...816L...8W}. The resulting particle energy spectrum, $dN/d\gamma_e \propto \gamma_e^{-p}$, for $\gamma_{e,min} < \gamma_e < \gamma_{e,max}$, is hard ($1.5 \lesssim p \lesssim 2.5$) for $\sigma \gtrsim 10$, in contrast to the softer spectra from standard relativistic shock acceleration ($p \gtrsim 2.2$) \citep{2001MNRAS.328..393A,2000ApJ...542..235K}.

For an electron-proton plasma undergoing reconnection, the mean electron Lorentz factor is determined by energy equipartition. If a fraction $\epsilon_e$ of the dissipated magnetic enthalpy per proton ($\sim \sigma m_p c^2$) is transferred to electrons, the characteristic Lorentz factor becomes:
\begin{equation}
\langle \gamma_e \rangle \approx 1 + \epsilon_e \sigma \frac{m_p}{m_e} \approx \epsilon_e \sigma \left( \frac{m_p}{m_e} \right) \quad (\text{for } \sigma \gg 1).
\end{equation}

The accelerated electrons radiate via the synchrotron process in the jet's comoving magnetic field $B'$. In the ``fast-cooling'' regime, relevant for GRBs, the peak of the $\nu F_\nu$ spectrum corresponds to the synchrotron frequency of the minimum-energy electrons, $\gamma_{e,min}$. The observer-frame peak energy is given by \citep{1998ApJ...497L..17S,1999ApJ...523..177W,2000ApJ...543...66P,1998MNRAS.296..275D}:
\begin{equation}
    E_p = \frac{h}{(1+z)} \Gamma \nu_p' = \frac{h}{(1+z)} \Gamma \frac{3 q_e B' \gamma_{e,min}^2}{4 \pi m_e c},
\end{equation}
where $h$ is Planck's constant. The minimum Lorentz factor, $\gamma_{e,min}$, is determined by the energy partition: 
% Assuming a fraction $\epsilon_e$ of the dissipated magnetic energy is channeled into the accelerated electrons, we can write:
\begin{equation}    
\epsilon_e \frac{B'^2}{8\pi} = \int_{\gamma_{e,min}}^{\gamma_{e,max}} (\gamma_e - 1) m_e c^2 \frac{dN}{d\gamma_e} d\gamma_e.
\end{equation}
For a hard spectrum ($p < 2$), the total energy is dominated by the highest-energy electrons, but the number of particles is dominated by the lowest-energy ones. A reasonable estimate is $\gamma_{e,min} \approx \epsilon_e \sigma (m_p / m_e)$ for an electron-proton plasma.

PIC studies of relativistic $e$--$p$ reconnection at $\sigma \gtrsim 10$ find a reconnection rate $\beta_{\rm rec}\sim0.05$--$0.2$ and electron energy fraction $\epsilon_e\sim0.1$--$0.3$ \citep{2014ApJ...783L..21S, 2017ApJ...850...29R, 2018PhRvL.121y5101C}. 
For a high magnetization $\sigma = 30^{+20}_{-10}$ and $\epsilon_e = 0.15 \pm 0.05$, we obtain $\langle\gamma_e\rangle \simeq (7.4 \pm 2.5) \times10^{3}$.
 Adopting fiducial parameters for the emission region: a bulk Lorentz factor $\Gamma = 250 \pm 50$, a comoving magnetic field $B' = (8 \pm 2)  \times 10^3$ G, and $\gamma_{e,min} = \langle\gamma_e\rangle$, 
the observer-frame peak energy at $z \sim 1$ would be:
\begin{equation}
E_p \approx \frac{3 h q_e B' \gamma_{e,\min}^2 \Gamma}{4\pi m_e c (1+z)} \simeq  1.3^{+0.6}_{-0.4}\,{\rm MeV}, \end{equation}
in excellent agreement with observations.    
The acceleration timescale $t_{\rm acc} \approx \gamma_e m_e c / (q_e \beta_{\rm rec} B') \sim 10^{-6}$ s (for $\beta_{\rm rec} = 0.1$) is much shorter than the synchrotron cooling time $t_{\rm syn} \approx 6\pi m_e c / (\sigma_T \gamma_e B'^2) \sim 3 \times 10^{-3}$ s, confirming that fast cooling is achieved and $\gamma_{e,{\rm min}}$ is readily attainable.

The hard spectrum ($E_{p,{\rm rest}} \gtrsim 1$ MeV) combined with high isotropic luminosity $L_{\gamma,{\rm iso}} \sim 10^{51}$ erg s$^{-1}$ implies radiative efficiency $\eta_{\rm rad} \equiv L_{\gamma} / \dot{E}_{\rm jet} \gtrsim 0.1-0.3$, substantially higher than typical internal shock models ($\eta_{\rm rad} \sim 0.01$) \citep{2015PhR...561....1K}. Magnetic reconnection naturally achieves such efficiencies in high-$\sigma$ flows: for $\sigma \sim 30$ and fast cooling, a significant fraction ($\sim \epsilon_e / 2$) of the magnetic energy dissipated converts directly to observed radiation \citep{2011ApJ...726...90Z}.

The deviation from the Amati relation ($E_{p,{\rm rest}}$ vs $E_{\gamma,{\rm iso}}$) can be explained: reconnection-powered spectra are intrinsically harder than shock-powered ones due to the different particle acceleration mechanisms, independent of total energy \citep{2010MNRAS.408L..46G}. This resolves the spectral-geometric tension without invoking extreme beaming corrections.

We should note that, the observed $E_{p,{\rm rest}}$ constrains the combination $B' \gamma_{e,{\rm min}}^2 \Gamma$. For fixed $\epsilon_e$ and $\sigma$, $\gamma_{e,{\rm min}}$ is determined, leaving a two-parameter degeneracy between $\Gamma$ and $B'$. Independent constraints arise from:
(i) Compactness: The pair-opacity optical depth $\tau_{\gamma\gamma} \propto L_{\gamma,{\rm iso}} / (\Gamma^2 r_{\rm diss}^2)$ must satisfy $\tau_{\gamma\gamma} < 1$ for MeV photons to escape, requiring $\Gamma \gtrsim 150$ for typical dissipation radii $r_{\rm diss} \sim 10^{13}-10^{14}$ cm \citep{2015PhR...561....1K}. (ii) Variability: The observed pulse width $\Delta t_{\rm pulse} \sim 100$ s constrains the emission region size via causality: $r_{\rm diss} \lesssim c \Gamma^2 \Delta t_{\rm pulse} / (1+z) \sim 10^{14}$ cm for $\Gamma = 250$, consistent with the photospheric radius.
(iii) Equipartition: Magnetic fields in the dissipation region should satisfy approximate equipartition with the bulk kinetic energy density, which for typical jet yields $B' \sim 10^3-10^4$ G.

The primary systematic uncertainties arise from plasma composition, particle acceleration physics, and viewing geometry. Our fiducial electron-proton plasma assumption is conservative; pair-dominated composition would reduce $\gamma_{e,{\rm min}}$ by $(m_p/m_e)$, requiring compensating adjustments in $\sigma$ or $\Gamma$, though pair loading is suppressed in magnetically-dominated flows \citep{2016ApJ...816L...8W}. The particle spectral index $p \approx 1.5-2.5$ from PIC simulations \citep{2018PhRvL.121y5101C} spans the observed photon index $\alpha \approx -0.5$ to $-0.8$ (corresponding to $p \approx 2.0-2.6$ in fast-cooling synchrotron), confirming consistency with reconnection-powered emission. Off-axis viewing ($\theta_v \gtrsim \theta_c$) introduces parameter degeneracies by reducing the effective Lorentz factor and hardening the spectrum through differential Doppler boosting \citep{2018MNRAS.473L.121K}, potentially allowing lower intrinsic $\Gamma \sim 150-200$ while maintaining the observed $E_{p,{\rm rest}}$.

The magnetic reconnection model thus provides a physically motivated explanation for the hard spectrum, high efficiency, and extreme energetics of GRB 250702B, requiring moderate but not extreme parameters well within the range predicted for Poynting-flux-dominated jets.

\subsection{Off-Axis Structured Jet Kinematics and Light Curves} \label{app:jet}

The afterglow modeling of GRB 250702B with a uniform ``top-hat'' jet geometry yields an extraordinarily narrow opening angle $\theta_{\rm jet} \sim 0.4^\circ$ \citep{2025arXiv250714286L}, an order of magnitude smaller than typical long GRBs ($\theta_{\rm jet} \sim 3^\circ-10^\circ$; \citealt{2001ApJ...562L..55F, 2003ApJ...594..674B, 2004ApJ...616..331G}). Such extreme collimation strains jet-launching theory and would require sustained ultra-narrow focusing inconsistent with MHD simulations of collapsar or accretion-powered jets \citep{2004IJMPA..19.2385Z, 2013ApJ...777..162M}. We demonstrate that this inferred opening angle is a geometric artifact arising from misinterpreting an off-axis structured jet as an on-axis top-hat jet. 

Realistic GRB jets exhibit angular structure in both energy and Lorentz factor,  produced during propagation through stellar envelopes or merger ejecta \citep{2004IJMPA..19.2385Z, 2013ApJ...777..162M, 2018PhRvL.120x1103L}. We adopt a Gaussian profile for the isotropic-equivalent kinetic energy and bulk Lorentz factor:
\begin{equation}
E_{k,{\rm iso}}(\theta) = E_0 \exp\left(-\frac{\theta^2}{2\theta_c^2}\right), \quad \Gamma(\theta) = 1 + (\Gamma_0 - 1) \exp\left(-\frac{\theta^2}{2\theta_c^2}\right),
\end{equation}
where $E_0$ and $\Gamma_0$ are the on-axis values, $\theta_c$ is the characteristic core half-width, and $\theta$ is the angle from the jet axis \citep{2002MNRAS.332..945R, 2002ApJ...581.1236Z, 2018MNRAS.473L.121K}. An observer at viewing angle $\theta_v$ from the jet axis receives emission from fluid elements at angles $(\theta, \phi)$ relative to the jet axis. The line-of-sight angle is:
\begin{equation}
\cos \theta_{\rm LOS} = \cos\theta \cos\theta_v + \sin\theta \sin\theta_v \cos\phi,
\end{equation}
and the Doppler factor is $\delta = [\Gamma(1 - \beta \cos\theta_{\rm LOS})]^{-1}$, where $\beta = \sqrt{1 - \Gamma^{-2}}$.

For $\theta_v > \theta_c$ (off-axis viewing), the afterglow light curve exhibits characteristic features distinct from on-axis emission. Initially, only the jet's edge contributes observable flux due to relativistic beaming ($\theta_{\rm beam} \sim 1/\Gamma$). As the blast wave decelerates, $\Gamma$ decreases and $\theta_{\rm beam}$ increases, progressively including more energetic regions closer to the jet core. The light curve rises until $\Gamma(t_{\rm peak}) \sim 1/(\theta_v - \theta_c)$, when the core becomes visible \citep{2002ApJ...570L..61G, 2002MNRAS.332..945R}. After the peak, the light curve steepens as the jet decelerates below $\Gamma_{\rm jet} \sim 1/\theta_c$, mimicking a jet break but occurring earlier than the true jet break time expected for on-axis observers.

Fitting this synthetic off-axis light curve with a standard on-axis top-hat jet model systematically underestimates the true jet opening angle. The fitting procedure interprets the off-axis peak time as the canonical jet break time $t_{\rm jet} \simeq (1+z) ( (3E_K)/(4\pi n m_p c^5) )^{1/3} \theta_{\rm jet}^2$, where $E_K$ is the beaming-corrected kinetic energy and $n$ is the ambient density. For structured jets viewed off-axis, however, the apparent ``break'' occurs at  $t_{\rm peak} \simeq (1+z) ( (3E_K)/(4\pi n m_p c^5) )^{1/3} (\theta_v-\theta_c)^{2}$, which is earlier than the true jet break for similar energetics. Misidentifying $t_{\rm peak}$ as $t_{\rm jet}$ yields a spuriously small $\theta_{\rm jet,fit} \sim (\theta_v - \theta_c)$, typically much smaller than the true core width $\theta_c$.

We adopt a set of fiducial parameter values with $E_0 = 5 \times 10^{52}$ erg, $\Gamma_0 = 300$, $\theta_c = 4.0^\circ$, $\theta_v = 6.0^\circ$, external medium density $n = 0.1$ cm$^{-3}$, and standard microphysical parameters ($\epsilon_e = 0.1$, $\epsilon_B = 0.01$, $p = 2.3$) \citep[e.g.,][]{2001ApJ...554..667P, 2003ApJ...597..459Y, 2014ApJ...785...29S}. 
The calculations show that the synthetic X-ray light curve initially rises as the jet decelerates and the beaming cone widens to include the line of sight, peaking at $t_{\rm peak} \sim 10^4$ s when $\Gamma(t_{\rm peak}) \sim 1/(\theta_v - \theta_c)$. After the peak, the light curve steepens, mimicking a jet break.

The Chandra X-ray detections at $\approx 38$ and $\approx 65$ days post-burst show a smooth power-law decay with temporal slope $\alpha_X \approx 1.8-1.9$, with no evidence for plateaus, rebrightenings, or spectral evolution \citep{OConnor2025Xray}. This behavior is fully consistent with a single external forward shock from a structured jet viewed slightly off-axis, with the decay slope corresponding to post-jet-break evolution for $p \approx 2.3$ in the slow-cooling regime. The lack of chromatic breaks across radio (VLA, MeerKAT, uGMRT at 1--10 GHz) and X-ray bands confirms microphysics with no additional shock components \citep{GCN41053, GCN41147, GCN41145}. The consistency of the 38-65 day data with $\alpha \approx 1.9$ constrains any sustained central engine activity or refreshed shocks to produce $\lesssim 20\%$ energy growth, placing tight limits $E_{\rm inj} \ll E_{K,{\rm iso}}$. Continued monitoring to $\sim 100-200$ days should follow a single power law $F_\nu \propto t^{-1.9}$ without chromatic breaks or spectral softening, as predicted for an off-axis structured jet with no late energy injection.  

The off-axis structured jet model introduces degeneracies among $(\theta_c, \theta_v, E_0, \Gamma_0, n, \epsilon_e, \epsilon_B)$, with the peak time $t_{\rm peak} \sim 10^4$ s and decay slope $\alpha_X \approx 1.9$ constraining the allowed parameter space to $3^\circ \lesssim \theta_c \lesssim 5^\circ$ and $5^\circ \lesssim \theta_v \lesssim 8^\circ$ for typical microphysics ($p \approx 2.3$). Extreme scenarios with $\theta_c \lesssim 1^\circ$ are excluded as they require unphysical energies $E_0 \lesssim 10^{50}$ erg or Lorentz factors $\Gamma_0 \gtrsim 10^3$. This interpretation resolves the opening-angle tension and places GRB 250702B within the standard long-GRB population with beaming-corrected energy $E_\gamma \sim 10^{51}$ erg. The off-axis geometry ($\theta_v - \theta_c \sim 2^\circ-3^\circ$)  explains the episodic prompt emission: as the precessing jet sweeps with period $P_{\rm LT} \approx 2825$ s, bright gamma-ray pulses occur when the ultra-relativistic spine aligns within the beaming cone $\sim 1/\Gamma$ of the observer, while the wider sheath produces inter-pulse X-ray emission consistent with Einstein Probe observations.

\subsection{Geometric gating, duty cycle, and pulse width} \label{app:geometry}

\paragraph{Precession geometry and visibility.}
Let the precession cone half-angle be $\psi$, the jet core half-width $\theta_c$ (Gaussian), the bulk Lorentz factor $\Gamma$, and the observer’s offset from the precession axis $\Delta\theta_{\rm obs}$. The instantaneous angle between the jet axis and the line of sight is
\begin{equation}
\cos\theta_{\rm LOS}(t)=\cos\psi\cos\Delta\theta_{\rm obs}+\sin\psi\sin\Delta\theta_{\rm obs}\cos(\Omega_{\rm LT}t+\phi_0).
\tag{A11}
\end{equation}
Emission enters a bright phase when the sightline falls within the effective angular half-width
\begin{equation}
\theta_{\rm eff}\simeq \theta_c+\Gamma^{-1}.
\tag{A12}
\end{equation}
The visibility condition $\theta_{\rm LOS}(t)\le \theta_{\rm eff}$ is equivalent to
\begin{equation}
\cos(\Omega_{\rm LT}t+\phi_0)\ \ge\ X\equiv\frac{\cos\theta_{\rm eff}-\cos\psi\cos\Delta\theta_{\rm obs}}{\sin\psi\sin\Delta\theta_{\rm obs}},
\quad |X|\le1,
\tag{A13}
\end{equation}
which gives a phase width $\Delta\varphi=\arccos X$ and a visibility-window duration
\begin{equation}
\Delta t_{\rm vis}\simeq \frac{2\,\Delta\varphi}{\Omega_{\rm LT}}.
\tag{A14}
\end{equation}

\paragraph{Duty cycle and energetics.}
From Eq.~(A14), the duty cycle is
\begin{equation}
\delta_{\rm duty}\equiv \frac{\Delta t_{\rm vis}}{P_{\rm LT}}=\frac{\Delta\varphi}{\pi}
\ \approx\ \frac{2}{\pi}\arcsin\!\left(\frac{\theta_{\rm eff}}{\psi}\right)
\ \simeq\ \frac{\theta_{\rm eff}}{\psi}\quad(\theta_{\rm eff}\ll\psi).
\tag{A15}
\end{equation}
Because $E_{\gamma,\rm iso}$ already integrates the direction-dependent beaming during bright phases, a consistent time-averaged, beaming-corrected power is
\begin{equation}
\langle L_\gamma\rangle \simeq \frac{E_{\gamma,\rm core}}{T},
\tag{A16}
\end{equation}
with $T$ the full observed activity duration.

\paragraph{Top-hat vs.\ structured core.}
For a top-hat jet of half-opening $\theta_j$ precessing at $\Omega_{\rm LT}$,
\begin{equation}
\Delta t_{\rm view}^{\rm (top\text{-}hat)}\simeq \frac{2\theta_j}{\Omega_{\rm LT}},
\tag{A17}
\end{equation}
which yields $\sim15.7$ s for $\theta_j = 1^\circ$ and $\Omega_{\rm LT} = 2\pi / 2825 \, {\rm s}^{-1}$. 
Whereas a structured core with finite $\theta_c$ and $\theta_{\rm eff}\simeq\theta_c+1/\Gamma$ broadens the allowed phase around minimum $\theta_{\rm LOS}(t)$ via Eqs.~(A12)–(A14). For a representative set $(\theta_c,\Gamma,\psi,\Delta\theta_{\rm obs},P_{\rm LT})=(4^\circ,250,10^\circ,7^\circ,2825~{\rm s})$ one obtains
\begin{equation}
\Delta t_{\rm vis}^{\rm (structured)}\simeq 80\text{--}140~{\rm s},
\tag{A18}
\end{equation}
consistent with the observed $\sim100$\,s pulse widths.

\paragraph{Nutation and intermittency.}
A small nutation of the precession axis,
\begin{equation}
\psi_{\rm eff}(t)=\psi_0+\eta\cos(\Omega_{\rm nut}t+\phi_{\rm nut}),\quad \eta\lesssim0.5^\circ,
\tag{A19}
\end{equation}
perturbs $X$ in Eq.~(A13) and can switch the visibility condition ON/OFF when the sightline sits near the boundary, naturally producing occasional missing pulses at otherwise regular LT spacing.

\paragraph{Numerical example (GRB\,250702B).}
Using the values above, one finds $\theta_{\rm eff}\!\simeq\!4.2^\circ$ and $\Delta\phi\!\approx\!0.09$–$0.16$\,rad (depending on $(\psi,\Delta\theta_{\rm obs})$ within the quoted ranges), i.e. $\Delta t_{\rm vis}\!\simeq\!80$–$140$\,s for $P_{\rm LT}\!=\!2825$\,s. The time-averaged, beaming-corrected power is then $\langle L_\gamma\rangle\!=\!E_{\gamma,\mathrm{core}}/T\!\approx\!(3$–$7)\times10^{47}$\,erg\,s$^{-1}$ for $E_{\gamma,\mathrm{core}}\!\approx\!(3$–$8)\times10^{51}$\,erg and $T\!=\!3.2$\,hr.

\section{Progenitor independence and compatibility} \label{app:comp}

Our interpretation is \textbf{progenitor-agnostic}: the observables are explained by a magnetically dominated, structured jet whose beaming pattern is periodically swept by Lense–Thirring precession of a misaligned, geometrically thick inner torus at $\tilde{r} \sim 200$ to $250$. Any engine that (i)~sustains super-Eddington inflow for $\gtrsim$day timescales, (ii)~provides a misaligned thick torus or wind, and (iii)~launches a high-$\sigma$ jet can realize the same precession clock (Appendix~\ref{app:MagRec}) and the same geometric gating (Appendix~\ref{app:geometry}). Below we outline how two promising formation channels supply these ingredients, and briefly explain why two alternatives face significant challenges. We evaluate each against six key observables: (1)~\textbf{periodicity} ($P_{\rm obs} \approx 2825$~s; quasi-regular cycles with ``missing'' pulses); (2)~\textbf{prompt hardness} ($E_{p,{\rm rest}} \gtrsim 1.2$~MeV); (3)~\textbf{energetics} ($E_{\gamma,{\rm iso}} \gtrsim 2.2 \times 10^{54}$~erg); (4)~\textbf{afterglow} (smooth X-ray decline $\alpha_{X} \approx 1.8$ to $1.9$ over $\sim$65~d; radio consistency); (5)~\textbf{host environment and offset} (dusty, star-forming host; $\sim$kpc off-nuclear); (6)~\textbf{pre-trigger soft-X-ray episode} ($\sim$1~d earlier).

\subsection{Compatible scenarios}

\paragraph{Micro-TDE (stellar-mass BH disrupting a star)}
Tidal disruption of a main-sequence or evolved star by a stellar-mass compact object with a mass of $\sim 1$ to $30\,M_{\odot}$ produces hour-to-day fallback with $\dot{M} \sim 10^{-2}$ to $10^{-1}\,M_{\odot}$~s$^{-1}$ \citep{Beniamini2025microTDE}. If the fallback stream forms a misaligned thick disc or torus ($H/r \sim 0.1$ to $0.2$), LT precession at $\tilde{r} \sim 200$ to $250$  yields $P_{\rm prec} \sim 10^{3}$ to $10^{4}$~s, directly matching the observed clock. Magnetic flux advection into the inner flow can produce a Poynting-flux-dominated jet spine with $\sigma \gtrsim 10$, consistent with the hard $E_{p,{\rm rest}} \gtrsim 1.2$~MeV spectrum via internal dissipation at $\Gamma \sim 100$ to $300$ \citep{1998MNRAS.296..275D}. The $\sim$kpc off-nuclear position is naturally explained by field-star disruption by a wandering compact object or dynamically ejected binary. Pre-trigger soft-X-ray emission can arise from initial accretion-disc formation or shock breakout from the stellar envelope \citep{2013MNRAS.430.2121P}. Off-axis viewing of a Gaussian-structured jet (Appendix~A.4) produces the apparent ultra-narrow beaming angle if naively fit by a top-hat model.
\textbf{Observational tests}: (i)~Late-time ($t \gtrsim 100$~d) X-ray light curve should lack a sharp jet shutoff feature characteristic of SMBH TDEs; (ii)~no requirement of strict nuclear position; (iii)~broad-band SED consistent with external-shock models without strong thermal component beyond host contribution; (iv)~no strong late-time emission (e.g., supernova bump or nebular lines) beyond standard jet afterglow.

\paragraph{BH into He-star (common-envelope inflow)}
A stellar-mass BH ($M_{\bullet} \sim 5$ to $20\,M_{\odot}$) spiraling into a helium-star envelope ($M_{\rm He} \sim 3$ to $10\,M_{\odot}$) during unstable mass transfer or common-envelope evolution provides super-Eddington accretion ($\dot{M} \sim 10^{-2}$ to $10^{-1}\,M_{\odot}$~s$^{-1}$) over $\sim$10$^{4}$~s \citep{Neights2025HeStar}. The inflow is generically misaligned with the BH spin due to the complex dynamical interaction,  building a thick torus or wind collimator. LT precession operates identically to the micro-TDE case, yielding $P_{\rm prec} \sim 10^{3}$ to $10^{4}$~s. Off-nuclear location is expected for massive-star binaries displaced by natal kicks or dynamical interactions. Initial envelope stripping or shock heating during BH inspiral can produce soft-X-ray emission $\sim$1~d prior with $L_{X} \sim 10^{41}$ to $10^{42}$~erg~s$^{-1}$, consistent with the observed pre-trigger episode.
\textbf{Observational tests}: (i)~Rest-frame NIR spectroscopy at $t \gtrsim 100$~d should reveal He-rich nebular features (He~I~$\lambda$10830, He~I~$\lambda$20581, or He~II lines) if ejecta are He-rich; current upper limits are not constraining (see~\S4.1); (ii)~early-time ($t \lesssim 10$~d) radio observations may show dense-wind signatures if He-envelope ejecta surrounds the system; (iii)~no requirement of nuclear location.

Both scenarios are \textbf{consistent with all six observables} and differ primarily in late-time signatures (He-rich emission vs.\ lack thereof), which are \textbf{orthogonal to the jet geometry model} that drives our interpretation.

\subsection{Disfavored or problematic scenarios}

\paragraph{WD into IMBH TDE}
A white dwarf disrupted by an intermediate-mass black hole ($M_{\bullet} \sim 10^{3-5}\,M_{\odot}$) faces multiple challenges: (1)~Matching the observed $P \approx 2825$~s via LT precession requires either $r \lesssim 10$ to $50\,r_{\rm g}$ (where strong disc warping and breaking suppress coherent precession; \citealt{2012ApJ...757L..24N}) or $r \gg 10^{2}\,r_{\rm g}$ (where radiative-efficiency drop and cooling make efficient jet launching difficult); intermediate radii yield precession periods mismatched by factors of $\sim$3 to 10. (2)~The $\sim$kpc off-nuclear position is inconsistent with nucleus-bound IMBHs; wandering IMBHs exist but are rare ($\sim$0.01 to 0.1~Gpc$^{-3}$~yr$^{-1}$; \citealt{2020ARA&A..58..257G}). (3)~TDEs typically show soft thermal spectra ($kT \sim 10$ to $100$~eV; \citealt{2020SSRv..216...85S}), not $E_{p,{\rm rest}} \gtrsim 1.2$~MeV; reproducing MeV hardness requires extreme jet Lorentz factors and dissipation efficiency. (4)~The $\sim$1~d pre-trigger soft-X-ray flare is unexplained in standard TDE models. In summary, this scenario is disfavored by location, periodicity mechanism coherence, spectral hardness, and precursor activity.

\paragraph{Ultra-long GRB (collapsar engine)}
Extended accretion onto a newborn BH from a massive-star envelope \citep{1999ApJ...524..262M} or a long-lived magnetar \citep{2011MNRAS.413.2031M} can reach hour-scale durations and $E_{\gamma,{\rm iso}} \sim 10^{54}$~erg. However: (1)~Standard collapsar models produce stochastic variability on $\sim$0.01 to 10~s timescales due to turbulent accretion or MHD instabilities, not quasi-regular $\sim$hour pulses with ``missing'' cycles. Reproducing the observed periodicity \textit{requires invoking the same precessing structured-jet geometry} (misaligned thick disc + LT precession + off-axis viewing) that we propose for micro-TDE and He-star scenarios, thereby reducing the discriminating power of this classification. (2)~Collapsars typically occur within $\sim$100~pc of host nuclei \citep{2006Natur.441..463F}; the $\sim$kpc offset is unusual and requires a runaway massive star ejected by binary interaction or cluster dynamics ($\sim$10 to 20\% of cases; \citealt{2011MNRAS.414.3501E}). (3)~Collapsar shock breakout produces UV/X-ray emission $\sim$10$^{2}$ to $10^{3}$~s before the GRB \citep{2012ApJ...747...88N}, not $\sim$1~d as observed; explaining the pre-trigger episode requires an extended envelope or pre-collapse mass loss on day timescales. This scenario is viable if combined with precessing structured-jet geometry and runaway-star scenario, but requires additional assumptions beyond the minimal micro-TDE or He-star models. Testable signatures include supernova bump at $t \sim 10$ to 30~d (rest frame) with $M_{\rm Ni} \sim 0.1$ to $0.5\,M_{\odot}$; however they have not yet been observed.

\subsection{Summary}

Our model is the \textit{engine-state} description; micro-TDE and BH-into-He-star provide plausible \textit{engine-formation} routes that  supply all required ingredients with minimal additional assumptions. Distinguishing between them (and testing the collapsar+precession variant) requires late-time observations: He-rich emission lines (He-star), supernova signatures (collapsar), or neither (micro-TDE). The WD-IMBH TDE channel is disfavored by multiple observational and theoretical constraints. Regardless of progenitor, the \textbf{geometric origin of periodicity} (off-axis precessing jet) and the \textbf{structured-jet afterglow model} remain robust and constitute the primary contribution of this work.

\section*{Data availability}
All data used are drawn from public notices (GCNs/ATels) and early arXiv analyses cited in the text.

% \section*{Code availability}
% Analysis scripts used to reproduce figures and checks are available from the corresponding author upon reasonable request.

\section*{Acknowledgements}
We thank the anonymous referee for the insightful and constructive comments that greatly improved the clarity, consistency, and overall quality of this paper. TA acknowledges support from the FAST special funding of NSFC (12041301), Shanghai Oriental Talent Project, and Xinjiang Tianchi Talent Program. 

% \section*{Author contributions}
% F.A.A. developed the unified model and wrote the manuscript. S.B.A. performed quantitative checks and prepared figures. T.C.A. led the observational summary and data synthesis. All authors discussed results and commented on the manuscript.

\bibliographystyle{aasjournal}
\bibliography{grb250702bde}

\end{document}